\documentstyle[twoside,psfig]{article}

\catcode`\@=11
\long\def\@makefntext#1{
\protect\noindent \hbox to 3.2pt {\hskip-.9pt  
$^{{\eightrm\@thefnmark}}$\hfil}#1\hfill}		

\def\thefootnote{\fnsymbol{footnote}}
\def\@makefnmark{\hbox to 0pt{$^{\@thefnmark}$\hss}}	
	
\def\ps@myheadings{\let\@mkboth\@gobbletwo
\def\@oddhead{\hbox{}
\rightmark\hfil\eightrm\thepage}   
\def\@oddfoot{}\def\@evenhead{\eightrm\thepage\hfil
\leftmark\hbox{}}\def\@evenfoot{}
\def\sectionmark##1{}\def\subsectionmark##1{}}

\oddsidemargin=\evensidemargin
\renewcommand{\thefootnote}{\fnsymbol{footnote}}

\newcounter{sectionc}\newcounter{subsectionc}\newcounter{subsubsectionc}
\renewcommand{\section}[1] {\vspace{12pt}\addtocounter{sectionc}{1} 
\setcounter{subsectionc}{0}\setcounter{subsubsectionc}{0}\noindent 
	{\tenbf\thesectionc. #1}\par\vspace{5pt}}
\renewcommand{\subsection}[1] {\vspace{12pt}\addtocounter{subsectionc}{1} 
	\setcounter{subsubsectionc}{0}\noindent 
	{\bf\thesectionc.\thesubsectionc. {\kern1pt \bfit #1}}\par\vspace{5pt}}
\renewcommand{\subsubsection}[1] {\vspace{12pt}\addtocounter{subsubsectionc}{1}
	\noindent{\tenrm\thesectionc.\thesubsectionc.\thesubsubsectionc.
	{\kern1pt \tenit #1}}\par\vspace{5pt}}
\newcommand{\nonumsection}[1] {\vspace{12pt}\noindent{\tenbf #1}
	\par\vspace{5pt}}

\newcounter{appendixc}
\newcounter{subappendixc}[appendixc]
\newcounter{subsubappendixc}[subappendixc]
\renewcommand{\thesubappendixc}{\Alph{appendixc}.\arabic{subappendixc}}
\renewcommand{\thesubsubappendixc}
	{\Alph{appendixc}.\arabic{subappendixc}.\arabic{subsubappendixc}}

\renewcommand{\appendix}[1] {\vspace{12pt}
        \refstepcounter{appendixc}
        \setcounter{figure}{0}
        \setcounter{table}{0}
        \setcounter{lemma}{0}
        \setcounter{theorem}{0}
        \setcounter{corollary}{0}
        \setcounter{definition}{0}
        \setcounter{equation}{0}
        \renewcommand{\thefigure}{\Alph{appendixc}.\arabic{figure}}
        \renewcommand{\thetable}{\Alph{appendixc}.\arabic{table}}
        \renewcommand{\theappendixc}{\Alph{appendixc}}
        \renewcommand{\thelemma}{\Alph{appendixc}.\arabic{lemma}}
        \renewcommand{\thetheorem}{\Alph{appendixc}.\arabic{theorem}}
        \renewcommand{\thedefinition}{\Alph{appendixc}.\arabic{definition}}
        \renewcommand{\thecorollary}{\Alph{appendixc}.\arabic{corollary}}
        \renewcommand{\theequation}{\Alph{appendixc}.\arabic{equation}}
        \noindent{\tenbf Appendix \theappendixc #1}\par\vspace{5pt}}
\newcommand{\subappendix}[1] {\vspace{12pt}
        \refstepcounter{subappendixc}
        \noindent{\bf Appendix \thesubappendixc. {\kern1pt \bfit #1}}
	\par\vspace{5pt}}
\newcommand{\subsubappendix}[1] {\vspace{12pt}
        \refstepcounter{subsubappendixc}
        \noindent{\rm Appendix \thesubsubappendixc. {\kern1pt \tenit #1}}
	\par\vspace{5pt}}

\newcommand{\textlineskip}{\baselineskip=13pt}
\newcommand{\smalllineskip}{\baselineskip=10pt}

\def\eightcirc{
\begin{picture}(0,0)
\put(4.4,1.8){\circle{6.5}}
\end{picture}}
\def\eightcopyright{\eightcirc\kern2.7pt\hbox{\eightrm c}} 

\newcommand{\copyrightheading}[1]
	{\vspace*{-2.5cm}\smalllineskip{\flushleft
	{\footnotesize International Journal of Modern Physics A #1}\\
	{\footnotesize $\eightcopyright$\, World Scientific Publishing
	 Company}\\
	 }}

\def\abstracts#1#2#3{{
	\centering{\begin{minipage}{4.5in}\footnotesize\baselineskip=10pt
	\parindent=0pt #1\par 
	\parindent=15pt #2\par
	\parindent=15pt #3
	\end{minipage}}\par}} 

\newcommand{\bibit}{\nineit}

\renewenvironment{thebibliography}[1]
	{\frenchspacing
	 \ninerm\baselineskip=11pt
	 \begin{list}{\arabic{enumi}.}
	{\usecounter{enumi}\setlength{\parsep}{0pt}
	 \setlength{\leftmargin 12.7pt}{\rightmargin 0pt} 
	 \setlength{\itemsep}{0pt} \settowidth
	{\labelwidth}{#1.}\sloppy}}{\end{list}}

\newcounter{itemlistc}
\newcounter{romanlistc}
\newcounter{alphlistc}
\newcounter{arabiclistc}

\newcommand{\fcaption}[1]{
        \refstepcounter{figure}
        \setbox\@tempboxa = \hbox{\footnotesize Fig.~\thefigure. #1}
        \ifdim \wd\@tempboxa > 5in
           {\begin{center}
        \parbox{5in}{\footnotesize\smalllineskip Fig.~\thefigure. #1}
            \end{center}}
        \else
             {\begin{center}
             {\footnotesize Fig.~\thefigure. #1}
              \end{center}}
        \fi}

\newcommand{\tcaption}[1]{
        \refstepcounter{table}
        \setbox\@tempboxa = \hbox{\footnotesize Table~\thetable. #1}
        \ifdim \wd\@tempboxa > 5in
           {\begin{center}
        \parbox{5in}{\footnotesize\smalllineskip Table~\thetable. #1}
            \end{center}}
        \else
             {\begin{center}
             {\footnotesize Table~\thetable. #1}
              \end{center}}
        \fi}

\def\@citex[#1]#2{\if@filesw\immediate\write\@auxout
	{\string\citation{#2}}\fi
\def\@citea{}\@cite{\@for\@citeb:=#2\do
	{\@citea\def\@citea{,}\@ifundefined
	{b@\@citeb}{{\bf ?}\@warning
	{Citation `\@citeb' on page \thepage \space undefined}}
	{\csname b@\@citeb\endcsname}}}{#1}}

\newif\if@cghi
\def\cite{\@cghitrue\@ifnextchar [{\@tempswatrue
	\@citex}{\@tempswafalse\@citex[]}}
\def\citelow{\@cghifalse\@ifnextchar [{\@tempswatrue
	\@citex}{\@tempswafalse\@citex[]}}
\def\@cite#1#2{{$\null^{#1}$\if@tempswa\typeout
	{IJCGA warning: optional citation argument 
	ignored: `#2'} \fi}}

\def\pmb#1{\setbox0=\hbox{#1}
	\kern-.025em\copy0\kern-\wd0
	\kern.05em\copy0\kern-\wd0
	\kern-.025em\raise.0433em\box0}

\def\fnt#1#2{\footnotetext{\kern-.3em
	{$^{\mbox{\scriptsize #1}}$}{#2}}}

\def\thefootnote{\fnsymbol{footnote}}
\def\@makefnmark{\hbox to 0pt{$^{\@thefnmark}$\hss}}	
	
\def\ps@myheadings{%
    \let\@oddfoot\@empty\let\@evenfoot\@empty
    \def\@evenhead{\slshape\leftmark\hfil}
    \def\@oddhead{\hfil{\slshape\rightmark}}
    \let\@mkboth\@gobbletwo
    \let\sectionmark\@gobble
    \let\subsectionmark\@gobble
    }
\font\tenrm=cmr10
\font\tenit=cmti10 
\font\tenbf=cmbx10
\font\bfit=cmbxti10 at 10pt
\font\ninerm=cmr9
\font\nineit=cmti9

\font\eightrm=cmr8

\def\qed{\hbox{${\vcenter{\vbox{			
   \hrule height 0.4pt\hbox{\vrule width 0.4pt height 6pt
   \kern5pt\vrule width 0.4pt}\hrule height 0.4pt}}}$}}

\renewcommand{\thefootnote}{\fnsymbol{footnote}}  

\begin{document}
\setlength{\textheight}{7.7truein}  

\thispagestyle{empty}

\markboth{\protect{\footnotesize\it The virtual black hole in 2d quantum gravity}}{\protect{\footnotesize\it The virtual black hole in 2d quantum gravity}}

\normalsize\textlineskip

\setcounter{page}{1}

\copyrightheading{}		

\vspace*{0.88truein}

\centerline{\bf THE VIRTUAL BLACK HOLE IN 2D QUANTUM GRAVITY}
\vspace*{0.035truein}
\centerline{\bf AND ITS RELEVANCE FOR THE S-MATRIX}
\vspace*{0.37truein}
\centerline{\footnotesize DANIEL GRUMILLER\footnote{e-mail: {\tt grumil@hep.itp.tuwien.ac.at}}}
\baselineskip=12pt
\centerline{\footnotesize\it Institut f{\"u}r Theoretische Physik, Technische Universit{\"a}t Wien, Wiedner Hauptstr. 8-10}
\baselineskip=10pt
\centerline{\footnotesize\it Vienna, A-1040,
Austria}
\vspace*{0.225truein}

\vspace*{0.21truein}
\abstracts{As shown recently 2d quantum gravity 
theories --- including spherically reduced Einstein-gravity ---  
after an exact path integral of its geometric part can be treated 
perturbatively in the loops of (scalar) matter. Obviously the 
classical mechanism of black hole formation should be contained 
in the tree approximation of the theory. This is shown to be the 
case for the scattering of two scalars through an intermediate 
state which by its effective black hole mass is identified as a 
``virtual black hole''. We discuss the lowest order tree vertex for minimally 
and non-minimally coupled scalars and find a non-trivial finite S-matrix
for gravitational s-wave scattering in the latter case.}{}{}



\vspace*{1pt}\textlineskip	
\section{Introduction to first order gravity in two dimensions}	
\vspace*{-0.5pt}
\noindent
Quantum gravity is beset with well-known conceptual problems. Probably the most
challenging one is the dual r{\^o}le of geometric variables as fields which at 
the same time determine the local and global properties of the manifold on 
which they act. Due to this fact and because gravity is perturbatively 
non-renormalizable it is desirable to use non-perturbative methods (however, 
cf. the article of M. Reuter ``Is Quantum Gravity Asymptotically Safe?'' in 
this volume). Unfortunately, in $d=4$ this is technically problematic. 

Therefore, models in $d=2$ are considered frequently in this context, most of 
which lack an important feature present in ordinary gravity: they contain no 
continuous physical degrees of freedom. One way to overcome this without 
leaving the comfortable realm of two dimensions is the inclusion of matter. 
We will restrict ourselves to the following class of actions:
\begin{equation}
L_{dil} = \int d^2x \sqrt{-g} \left[ X R - U(X)
\left(\nabla X\right)^2 + 2 V(X) + F(X) \left( \nabla \phi \right)^2 \right].
\label{dil}
\end{equation}
$R$ denotes the $2d$ scalar curvature, $X$ is the so-called dilaton field, 
$\nabla$ the covariant derivative containing the Levi-Civit{\'a} connection 
with respect to the metric $g_{\mu\nu}$ and $\phi$ is a massless scalar field. 
Adjusting the functions $U(X)$, $V(X)$ and $F(X)$ properly, one obtains a 
variety of interesting models, the most prominent among them being the CGHS 
model\cite{Callan:1992rs} and spherically reduced gravity\cite{Thomi:1984na}
(SRG). In this work we will focus on (originally fourdimensional) SRG. 

The action (\ref{dil}) is locally and globally equivalent to a first order 
one\cite{Ikeda:1993aj,Kummer:1995qv} depending on 
Cartan 1-forms $e^{\pm}$ (we denote light-cone indices with $\pm$) and 
$\omega$ (the abelian gauge structure of the two dimensional spin connection 
$\omega^a_{\quad b}=\varepsilon^a_{\quad b} \omega$ is
used explicitly), the dilaton field $X$, the auxiliary 
fields $X^{\pm}$ and the Klein-Gordon field $\phi$:
\begin{eqnarray}
L_{FO} &=& \int {\big [} X^+ (d - \omega) \wedge e^- + 
X^- (d + \omega) \wedge e^+ + \nonumber \\
&& + X d\wedge \omega - e^- \wedge e^+ {\cal V} - F(X) d\phi 
\wedge *d\phi {\big ]}. \label{Q20}
\end{eqnarray}  
Actually, this equivalence holds for general dilaton theories. In the present
spherically reduced case the ``potential'' in (\ref{Q20}) becomes 
${\cal V} = -1 - \frac{X^+X^-}{2X}$. The coupling function $F(X)$ will be 
chosen as a constant (minimally coupled case) or linear in $X$ (non-minimally 
coupled case with proper $s$-wave factor). The latter case arises if one 
considers a minimally coupled massless scalar field in $d=4$, the factor $X$ 
being a remnant of the $4d$ measure $\sqrt{-g^{(4)}}\propto|X|\sqrt{-g^{(2)}}$.
Note that $X$ is restricted to the half-line, which is why often alternatively 
the representation $X=\exp{(-2\Phi)}$ is used with $\Phi\in(-\infty,\infty)$.

\setcounter{footnote}{0}
\renewcommand{\thefootnote}{\alph{footnote}}

\section{Path integral quantization}
\noindent
The BRST analysis and path-integral quantization of models of the type 
(\ref{Q20}) has been discussed for the minimally 
coupled case\cite{Kummer:1998zs} as well as for the more complicated 
non-minimally coupled case\cite{Grumiller:2001ea}.
Using a convenient gauge for the Cartan variables (namely ``temporal 
gauge'': $\omega_0=0$, $e^+_0=0$ and $e^-_0=1$) an exact integration of 
all geometric variables yields a non-local and non-polynomial action depending 
solely on the scalar field and on eventual external sources. We focus on the 
lowest order tree diagrams, i.e. on the (non-local) $\phi^4$-vertices. Thus,
we use perturbation theory only in the matter sector, implicitly assuming that 
the typical energies of the scalar particles are small as compared to Planck 
energy. In addition to our local gauge fixing we had to impose certain boundary
conditions. We chose for simplicity conditions yielding asymptotically
Minkowski space-time.

For minimally coupled massless scalars we obtained a non-trivial vertex, but 
only a trivial result for the scattering amplitude: It was either divergent or 
-- if certain regularity conditions were imposed by hand -- it vanished (the 
black hole was ``plugged'' by these boundary conditions). For massive scalars 
we obtained a non-trivial result for the $S$-matrix\cite{Grumiller:2000ah}.

The more interesting non-minimally coupled case yielded two vertices even for 
massless particles (cf. fig. \ref{fig:1}).
\begin{figure}[htbp]
\vspace*{13pt}
\centerline{\psfig{file=bothphi.epsi,width=0.8\textwidth}} 
\vspace*{13pt}
\fcaption{Lowest order (non-local) tree vertices with outer legs.}
\label{fig:1}
\end{figure}
The explicit expressions for the quantities $V^{(4)}(x,y)_a$ and 
$V^{(4)}(x,y)_b$ contain simple polynomials multiplied by a
signum function depending of one coordinate pair (e.g. $x^0,y^0$) and they are 
local in the other pair. These properties are also present in the line element
discussed in the subsequent section\cite{Grumiller:2001ea,Fischer:2001vz}.

\newpage
\section{The virtual black hole}
\noindent
After obtaining a (perturbative or exact) solution for the matter degrees of 
freedom one can reconstruct the geometry. We did this for the lowest order 
tree-graph solution and found in both scenarios (minimal and non-minimal 
coupling) an effect which we called ``virtual black hole'' (VBH). We will 
briefly discuss this effect now.

In all theories of the type (\ref{Q20}) there exists a conserved 
quantity\cite{Kummer:1992rt}, even in the presence of 
matter\cite{Kummer:1995qv}. Its geometric part is essentially equivalent to 
the so-called ``mass aspect function'', a quantity widely used in numerical 
calculations to signal an apparent horizon\cite{Grumiller:1999rz}. It
has a discontinuity (a feature which is also present in the minimally coupled 
case) which is inherited by the effective line element\footnote{The
outgoing Bondi-Sachs form ($u=t-r$) of the line element is a direct 
consequence of the chosen gauge.}
\begin{equation}
(ds)^2=2drdu+\left(1-\frac{2m(r,u)}{r}-a(r,u)r\right)(du)^2,
\label{ds}
\end{equation}
with $m(r,u)=\delta(u-u_0)\theta(r_0-r)m_0(r_0)$ and 
$a(r,u)=\delta(u-u_0)\theta(r_0-r)a_0(r_0)$, $m_0$ and $a_0$ being smooth 
functions of $r_0$. The $\theta$-function is responsible for the 
discontinuity: For $r\to\infty$ neither the Schwarzschild term proportional to
$m$ nor the Rindler term proportional to $a$ is present.

This discontinuity effect in the mass-aspect function has been called 
``virtual black hole''. It is also reflected in the scalar curvature $R$, i.e.
it is not merely an artifact of an unsuitable coordinate system. However, this
VBH geometry should not be taken at face value, since the only observable 
in our context is the $S$-matrix.

\section{Scattering amplitude}
\noindent
Using asymptotic $s$-waves for the scalar field one can construct the Fock 
space and calculate the $S$ matrix element with ingoing modes $q,q'$ and
outgoing ones $k,k'$
\begin{equation}
T(q, q'; k, k') = -\frac{i\kappa\delta\left(k+k'-q-q'\right)}{2(4\pi)^4
|kk'qq'|^{3/2}} E^3 \tilde{T}
\label{RESULT}
\end{equation}
with the total energy $E=q+q'$, the coupling constant $\kappa=8\pi G_N$,
\begin{eqnarray}
\tilde{T} (q, q'; k, k') := \frac{1}{E^3}{\Bigg [}\Pi \ln{\frac{\Pi^2}{E^6}}
+ \frac{1} {\Pi} \sum_{p \in \left\{k,k',q,q'\right\}}p^2 \ln{\frac{p^2}{E^2}} 
\nonumber \\
\cdot {\Bigg (}3 kk'qq'-\frac{1}{2}
\sum_{r\neq p} \sum_{s \neq r,p}\left(r^2s^2\right){\Bigg )} {\Bigg ]},
\label{feynman}
\end{eqnarray}
and the momentum transfer function $\Pi = (k+k')(k-q)(k'-q)$. The interesting 
part of the scattering amplitude is encoded in the scale independent (!) 
factor $\tilde{T}$. This remarkably simple result has been discussed in more 
detail elsewhere\cite{Grumiller:2001ea,Fischer:2001vz}.

The idea that black holes must be considered in the $S$-matrix together
with elementary matter fields has been put forward some time 
ago\cite{'tHooft:1996tq}. Our approach has allowed for the first time
to derive (rather than suppose) the existence of the black hole states in the
quantum scattering matrix. So far, we were able to perform actual
computations in the first non-trivial order only. The next order
calculations which should yield an insight into the information paradox are in
progress.

\nonumsection{Acknowledgements}
\noindent
This work has been supported by project P-14650-TPH of the Austrian Science 
Foundation (FWF). I am grateful to my collaborators P. Fischer, W. Kummer and 
D. Vassilevich for many enlightening discussions and a careful reading of this
manuscript.

\nonumsection{References}

\eject

\end{document}